\documentclass{article}
\usepackage[utf8]{inputenc}
\usepackage{authblk}
\usepackage{setspace}
\usepackage[margin=1.25in]{geometry}
\usepackage{graphicx}
\graphicspath{ {./figures/} }
\usepackage{subcaption}
\usepackage{amsmath}
\usepackage{lineno}
\usepackage{hyperref}
\usepackage{xcolor}
\usepackage{soul}
\usepackage{siunitx}
\usepackage{wasysym}
\usepackage{amssymb}

\usepackage[
    backend=biber,
    style=numeric-comp,
    sorting=none,
    maxbibnames=3,
    giveninits=true
]{biblatex}
\title{A Broadband Squeezed Light Source for Table-Top Interferometry}
\date{}

\author[1,*]{Fabio Bergamin}
\author[1]{Nikitha Kuntimaddi}
\author[1]{Abhinav Patra}
\author[1]{Stephanie Montoya}
\author[2,3]{Moritz Mehmet}
\author[1]{Katherine Dooley}
\author[1]{Hartmut Grote}
\author[2,3]{Henning Vahlbruch}

\affil[1]{
School of Physics and Astronomy, Cardiff University, The Parade, CF24 3AA, United Kingdom}
\affil[2]{Max Planck Institute for Gravitational Physics (Albert Einstein Institute), D-30167 Hannover, Germany}
\affil[3]{Leibniz Universit\"at Hannover, D-30167 Hannover, Germany}
\affil[*]{bergaminf@cardiff.ac.uk}

\begin{document}

\maketitle

\begin{abstract}
We report on the characterisation of one of two broadband squeezed light sources developed for the Quantum Enhanced Space-Time (QUEST) experiment, using balanced homodyne detection. QUEST consists of a pair of co-located, table-top, power-recycled Michelson interferometers designed to probe stationary space-time fluctuations. The interferometers are designed to be shot-noise limited in the frequency range from 1 to 200\,MHz, and squeezed light will be employed with the goal to reduce the shot noise by 6\,dB at frequencies inside the linewidth of the optical parametric amplifier (OPA).\\
We directly observed up to 6.8\,dB of squeezing and maintained at least 3\,dB of squeezing across the full 100\,MHz measurement bandwidth. After accounting for the dark noise contribution, the inferred squeezing level increased to\,8.6 dB. Our squeezed light source is based on a hemilithic OPA with a 43.6\,mm round-trip optical length and a linewidth of 138\,MHz, making it the broadest-linewidth device to date among those suitable for long-term operation.
\end{abstract}

\section{Introduction}
Squeezed light refers to a class of non-classical states of the electromagnetic field in which the quantum noise in one field quadrature is reduced below the vacuum noise level, at the cost of increased noise in the conjugate quadrature \cite{schnabel2017squeezed}.\\
\indent The generation of squeezed states requires non-linear optical interactions, which can be realised through several physical processes, including optomechanical squeezing \cite{purdy2013strong}, the Kerr effect \cite{shelby1986broad}, four-wave mixing \cite{slusher1985observation}, and parametric down-conversion (PDC) \cite{wu1986generation}.
PDC can be implemented in different configurations: waveguide-based approaches \cite{kashiwazaki2023over, kashiwazaki2021fabrication, pysher2009broadband} offer the advantage of large bandwidth, whereas resonator-enhanced PDC \cite{wu1987squeezed} is notable for achieving the highest levels of squeezing \cite{vahlbruch2016detection}, albeit at the cost of reduced bandwidth. Resonator implementations can be broadly classified into two types: monolithic designs \cite{mehmet2010observation, ast2013high, tohermes2024directly}, which offer a compact structure with very small cavities but lack tunability in cavity length, and hemilithic \cite{vahlbruch2010geo,vahlbruch2016detection, mehmet2018high, lough2021first, heinze2022observation, Ast:12, meylahn2022squeezed, mehmet2011squeezed,schonbeck201713,darsow2021squeezed} or cavity-embedded crystal designs \cite{wilken2024broadband, meylahn2022squeezed, serikawa2016creation}, which allow for controllable cavity length.\\
\indent Squeezed light finds important applications in various fields, including quantum communication \cite{hillery2000quantum, ralph1999continuous} and quantum metrology. In particular, the use of squeezed states has become a standard technique in gravitational wave detectors \cite{tse2019quantum, acernese2019increasing,lough2021first}, which are based on Michelson interferometers. In such interferometers, quantum noise originates from vacuum fluctuations entering the dark port \cite{caves1981quantum}, and it can be reduced by replacing the coherent vacuum with a squeezed state \cite{dwyer2022squeezing}.\\
\indent The Quantum Enhanced Space-Time (QUEST) experiment \cite{vermeulen2021experiment, patra2024direct} is one of such precision measurement experiments that will make use of squeezed light. Located in Cardiff, UK, it consists of two co-located power-recycled Michelson interferometers with arm lengths of 1.84 m. The detector output is obtained by cross-correlating the signals from the two interferometers, a strategy that suppresses uncorrelated noise while enhancing common-mode signals.
The primary objective is to search for stationary space-time fluctuations—spacetime correlations arising from quantum gravity—but the instrument is also sensitive to signatures from dark matter and ultra-high frequency gravitational waves. In its final configuration, the detector will have a shot-noise-limited observation bandwidth from 1 to 200 MHz. To reduce the shot noise, squeezed vacuum states will be injected into the dark ports of both interferometers, using two independent squeezed light sources (SLSs).\\
\begin{figure}[h!]
\center{
\includegraphics[width=0.7\textwidth]{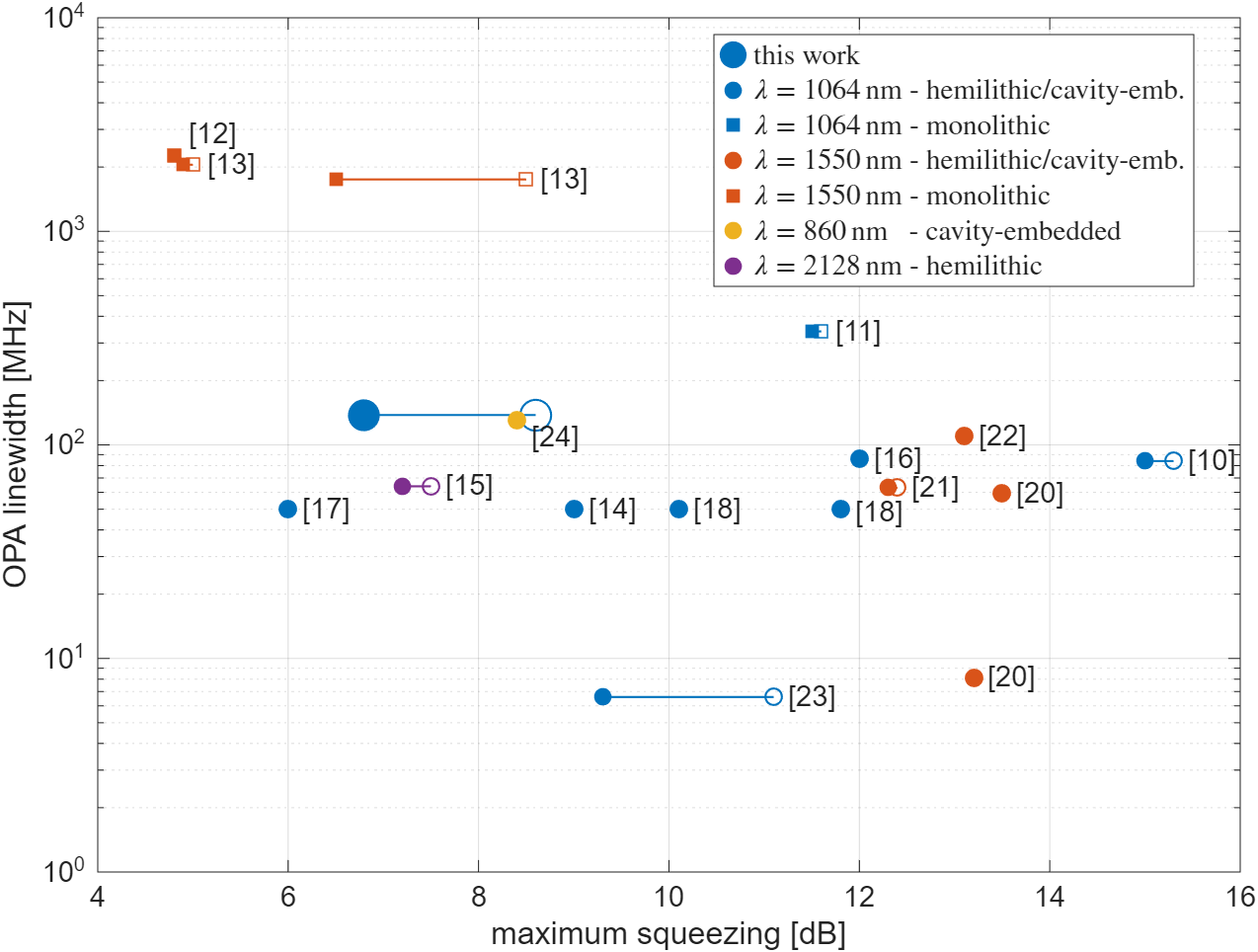}
\caption{\label{fig:previous_works} Summary of previous results. As shown in the plot, the SLS characterised in this work has the broadest linewidth among the OPAs with hemilithic or cavity-embedded design (hence suitable for long-term operation). The linewidth of our OPA (138\,MHz) is at least 6\,\% broader than the previous results \cite{serikawa2016creation}. $\Circle$, $\Square$: dark noise subtracted; $\CIRCLE$, $\blacksquare$: with dark noise.} }
\end{figure}
\indent In this paper, we present the characterisation of one of the two SLSs developed for the QUEST experiment. The source is based on the shortest hemilithic optical parametric amplifier (OPA) realised to date and produces the broadest-band squeezed light at 1064 nm reported in the context of interferometric applications. While even broader squeezing linewidths can be achieved using monolithic OPAs (see Fig.~\ref{fig:previous_works}), such devices are unsuitable for long-term integration with interferometers and are typically limited to bench-top measurements using balanced homodyne detection (BHD). This is because monolithic cavities require laser frequency tuning to maintain resonance, as active tuning of the crystal length via piezoelectric actuators is both limited in range and prone to mechanical distortions. However, in interferometric detectors, the laser frequency is typically fixed to resonate in other optical cavities (e.g., the power-recycling cavity in QUEST). As a result, the OPA cavity length must be tuneable to follow the laser frequency. For this reason, hemilithic OPAs offer a practical solution, and, to the best of our knowledge, our SLS achieves the broadest linewidth among those to be applied in interferometric measurements.\\
\indent The implementation of a BHD to measure the shot noise reduction provided by the SLS enabled us to characterise its performance in terms of linewidth and squeezing level. It also allowed us to test locking strategies for the readout phase control and to identify the dominant sources of optical loss, which is essential for minimising them in the final interferometer implementation.

\section{Squeezed Light Source}
\indent The SLS is a fully automated stand-alone system developed and constructed at the AEI Hannover, Germany. The system was built on a 
60\,cm x 60\,cm optical breadboard sealed with side panels and a top plate. 
The system does not require its own laser; when implemented into the QUEST experiment, each SLS will be supplied by a 470\,mW pick-off from the main laser feeding the corresponding interferometer.
Since this SLS is intended to be effective in the higher MHz range, the negative impact of technical laser noise at low frequencies is less critical than in sources optimised for audio-band squeezing \cite{vahlbruch2010geo,mehmet2020squeezed,tse2019quantum}. Therefore, the implementation of frequency-shifted coherent control fields \cite{vahlbruch2006coherent} is not required.\\
\begin{figure}[h!]
\center{
\includegraphics[width=1\textwidth]{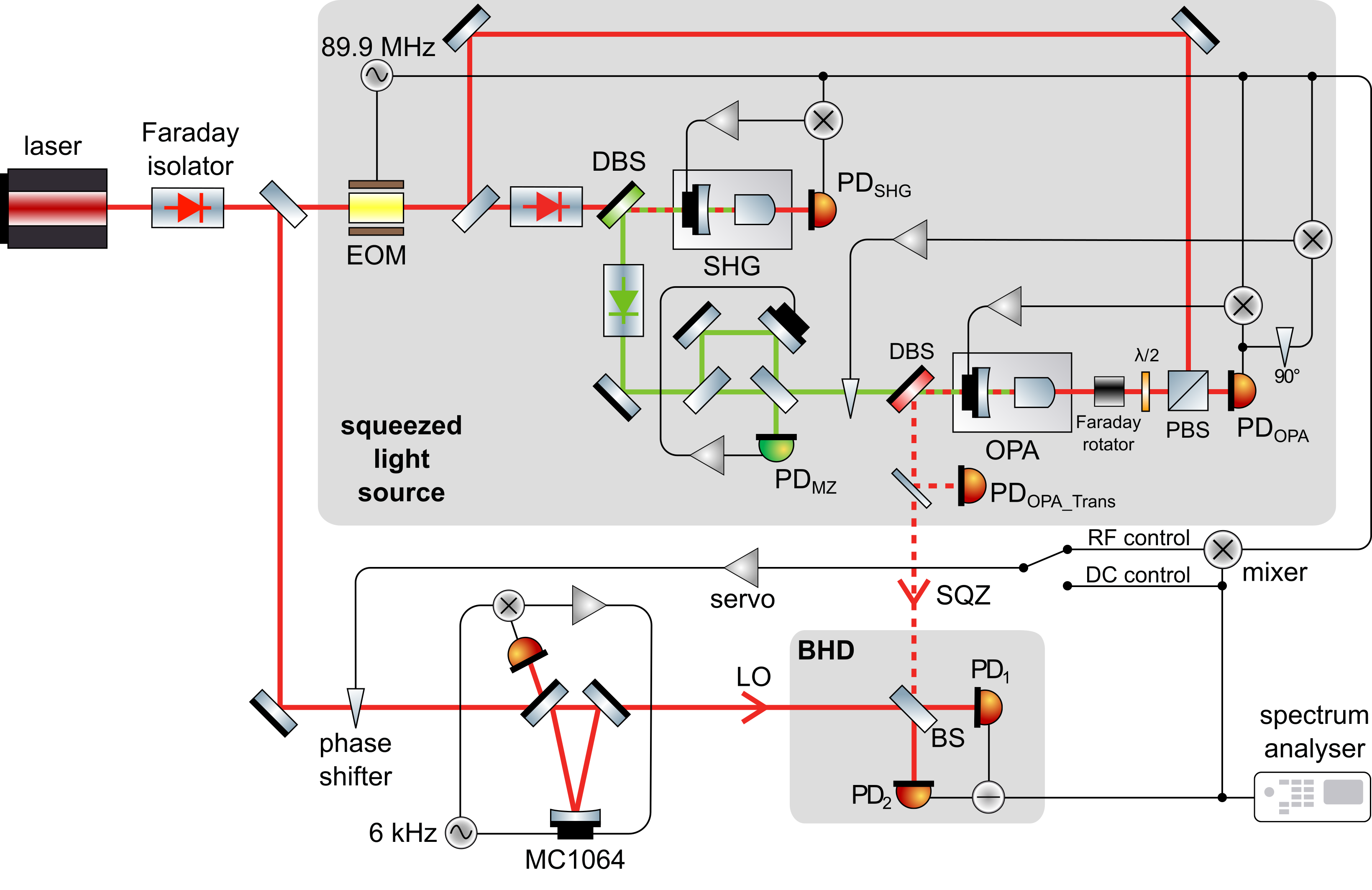}
\caption{\label{fig:squeezer} Layout of the SLS and of the BHD, including the control loops. BS: 50:50 beam splitter, DBS: dichroic beam splitter, EOM: electro-optic modulator, LO: local oscillator, MC: mode cleaner, MZ: Mach-Zehnder interferometer, OPA: optical parametric amplifier, PBS: polarising beam splitter, PD: photo-detector, SHG: second-harmonic generator, SQZ: squeezed light.}}
\end{figure}
\indent Fig.~\ref{fig:squeezer} shows a schematic of the SLS setup and of the detection scheme. The laser source was a 440 mW continuous-wave NPRO laser operating at a wavelength of 1064 nm.
An electro-optic modulator (EOM) at the input of the SLS phase-modulates the field at 89.9\,MHz. This modulation provides sidebands used for the longitudinal control of the optical cavities, the stabilisation of the pump phase, and the lock for the readout phase at the interferometer output (or at the BHD).\\
\indent A fraction (340\,mW) of the input infrared (IR) light was sent to a second-harmonic generator (SHG) to provide up to 240\,mW of light at 532\,nm. The SHG consists of a hemilithic cavity containing a plano-convex periodically poled potassium titanyl phosphate (PPKTP) crystal.
The SHG temperature was actively controlled to achieve quasi-phase-matching \cite{fejer2002quasi}. The cavity length was stabilised using a Pound–Drever–Hall (PDH) \cite{drever1983laser} error signal derived from a photo-detector (PD) in transmission ($\mathrm{PD_{SHG}}$ in Fig.~\ref{fig:squeezer}).\\
\indent A Mach-Zehnder (MZ) interferometer, operating at an offset from the bright fringe, is used to stabilise the power of the frequency-doubled field generated by the SHG. This frequency-doubled field acts as the pump, which drives the parametric (de-)amplification process in the optical parametric amplifier (OPA).
A power monitor in the pump path to the OPA ($\mathrm{PD_{MZ}}$ in Fig.~\ref{fig:squeezer}) serves as the sensor in the active feedback loop that controls the MZ interferometer.\\
\indent Similar to the SHG, the OPA is also constructed as a hemilithic cavity. The PPKTP crystal measures 9.3 x 2 x 1 $\mathrm{mm}^3$, its curved surface is high-reflection (HR) coated for both wavelengths, while the flat surface is anti-reflection coated for both wavelengths. The coupling mirror has a reflectivity of $R=88\,\%$ at 1064\,nm and $R < 2\,\%$ at 532\,nm, its radius of curvature is 9\,{mm} and is placed at 4.8\,{mm} from the flat surface of the crystal.  The round-trip optical length of $l_\mathrm{opt}=43.6\,\mathrm{mm}$ results in an FSR of 6.87\,GHz.
The OPA temperature was actively controlled for phase matching. About 5.1\,mW of the input laser was injected through the HR side of the OPA to generate the PDH error signal for longitudinal control, obtained from a PD in reflection of the cavity ($\mathrm{PD_{OPA}}$ in Fig.~\ref{fig:squeezer}). A PD  monitoring a 1\,\% pick-off in transmission of the OPA ($\mathrm{PD_{OPA\_Trans}}$ in Fig.~\ref{fig:squeezer}) is used to set the threshold for the OPA longitudinal lock. The bright IR field entering the OPA acts as a seed for the parametric amplification, resulting in the generation of a displaced squeezed state (or bright squeezing) rather than squeezed vacuum. The bright field exiting the OPA towards the BHD is used for alignment, locking on the anti-squeezing quadrature (see Sec.~\ref{sec:BHD}), and measuring the homodyne efficiency via fringe contrast (see Sec.~\ref{sec:imperfections}).\\
\indent The same PD used for the OPA longitudinal control is employed to obtain the error signal for the pump phase control by demodulating in quadrature with respect to the PDH demodulation phase \cite{hage2010purification, eberle2013realization}; the pump error signal thus obtained can lock the relative phase of the pump (532 nm) and the seed (1064 nm) in the OPA only in the configuration for which a de-amplification of the seed is achieved.\\
\indent The SLS features independent automatic locking of the SHG, MZ, OPA, and pump phase, with each sub-system acquiring lock within 1 second.

\section{Balanced Homodyne Detection}\label{sec:BHD}
\begin{figure}[h!]
\center{
\includegraphics[width=1\textwidth]{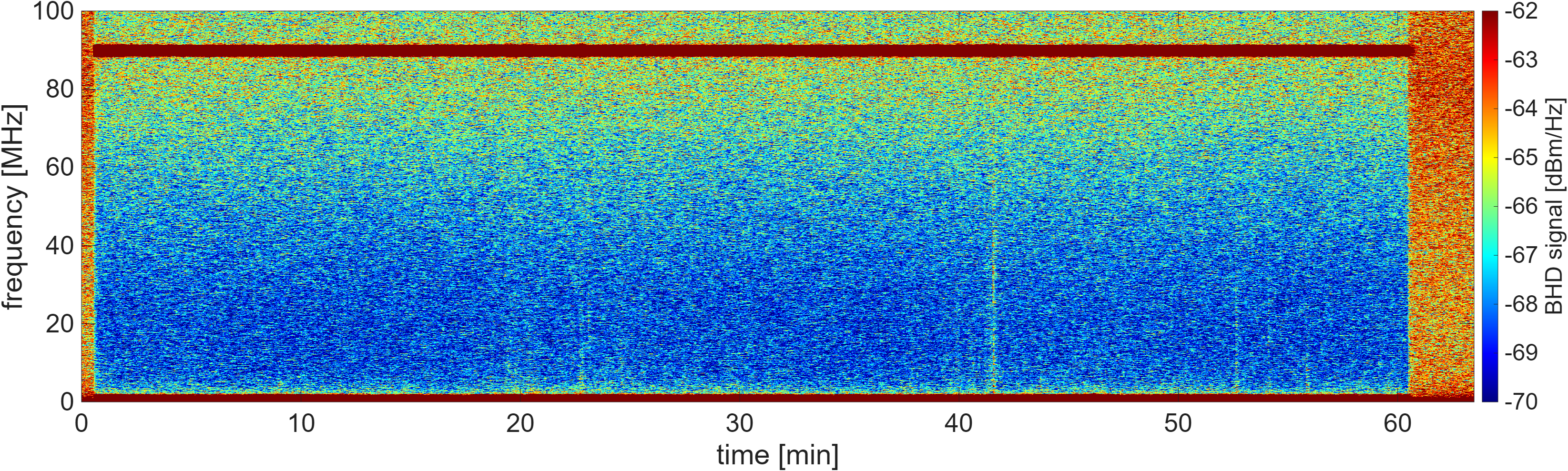}
\caption{\label{fig:spectrogram} Spectrogram of the BHD signal recorded during a one-hour continuous lock. A beam dump blocked the squeezed light path during the first 30 seconds and again after one hour. The spectrogram was obtained with a span of 100 MHz, RBW $=$ 1 MHz, sweep time of 1 s, and 5-trace moving average.}}
\end{figure}

The measurements of the coherent, squeezed, and anti-squeezed shot noise were performed using BHD. In the BHD setup, the signal and local oscillator (LO) fields are overlapped at a 50:50 beam splitter (BS). The two output beams are detected by PDs, and the BHD signal is obtained by subtracting the corresponding photocurrents.
The BHD allows measurement of the variance of any quadrature, depending on the phase $\Theta$ between the signal beam and the LO, which determines the quadrature being measured, according to
\begin{equation}
    \hat{X}(\Theta)=\hat{X}_A\cos(\Theta)+\hat{X}_P\sin(\Theta)\,,
\end{equation}
where $\hat{X}_A$ and $\hat{X}_P$ are the amplitude and phase quadrature operators, respectively.
In order to remove amplitude noise coupling of the LO to the BHD signal, the LO power should be much higher than the power of the signal field (the squeezed light). In our experiment, the power of the LO field was 17.2 mW, while that of the bright-alignment beam (BAB) was $\SI{31}{\micro\watt}$. The BAB refers to the bright IR field exiting the OPA in the absence of parametric amplification, i.e., when the pump field is blocked using a beam dump.\\
\indent To ensure high detection contrast, the LO was spatially cleaned using a mode cleaner (MC1064 in Fig.~\ref{fig:squeezer}). The mode cleaner was a triangular cavity \cite{griffiths2023enhancing} which was locked on resonance using a dither locking scheme \cite{buchler2001electro} with modulation frequency of 6\,kHz.
Another triangular cavity (not shown in Fig.~\ref{fig:squeezer}), identical in specification to the mode cleaner, was used as a reference cavity to align and mode-match the two beams at the BS. The cavity was placed in one of the BS's output ports and during normal operation, it was bypassed using a flip mirror.\\
\indent The two outputs of the BS were detected using high quantum efficiency InGaAs PDs, with an active area of  0.5 mm diameter. The PD electronics are custom-made to allow the detection of up to 50 mW optical power, with a bandwidth of 200 MHz. The linearity of the PD response was tested up to at least 200 MHz in a dedicated measurement.\\
The BHD signal was obtained by digitally subtracting the signals from the PDs using an FPGA (Moku:Pro) operating at a sampling rate of 312 MSa/s. Due to the limited speed of the digital subtraction process, the BHD measurement was only reliable up to the Nyquist frequency of 156 MHz. However, with the planned implementation of SLSs in the QUEST experiment, the signal will be digitised at 500 MSa/s using NI PXIe-5763 digitizers from National Instruments, enabling measurements across the full observation band of 200 MHz.\\
\begin{figure}[h!]
\center{
\includegraphics[width=.7\textwidth]{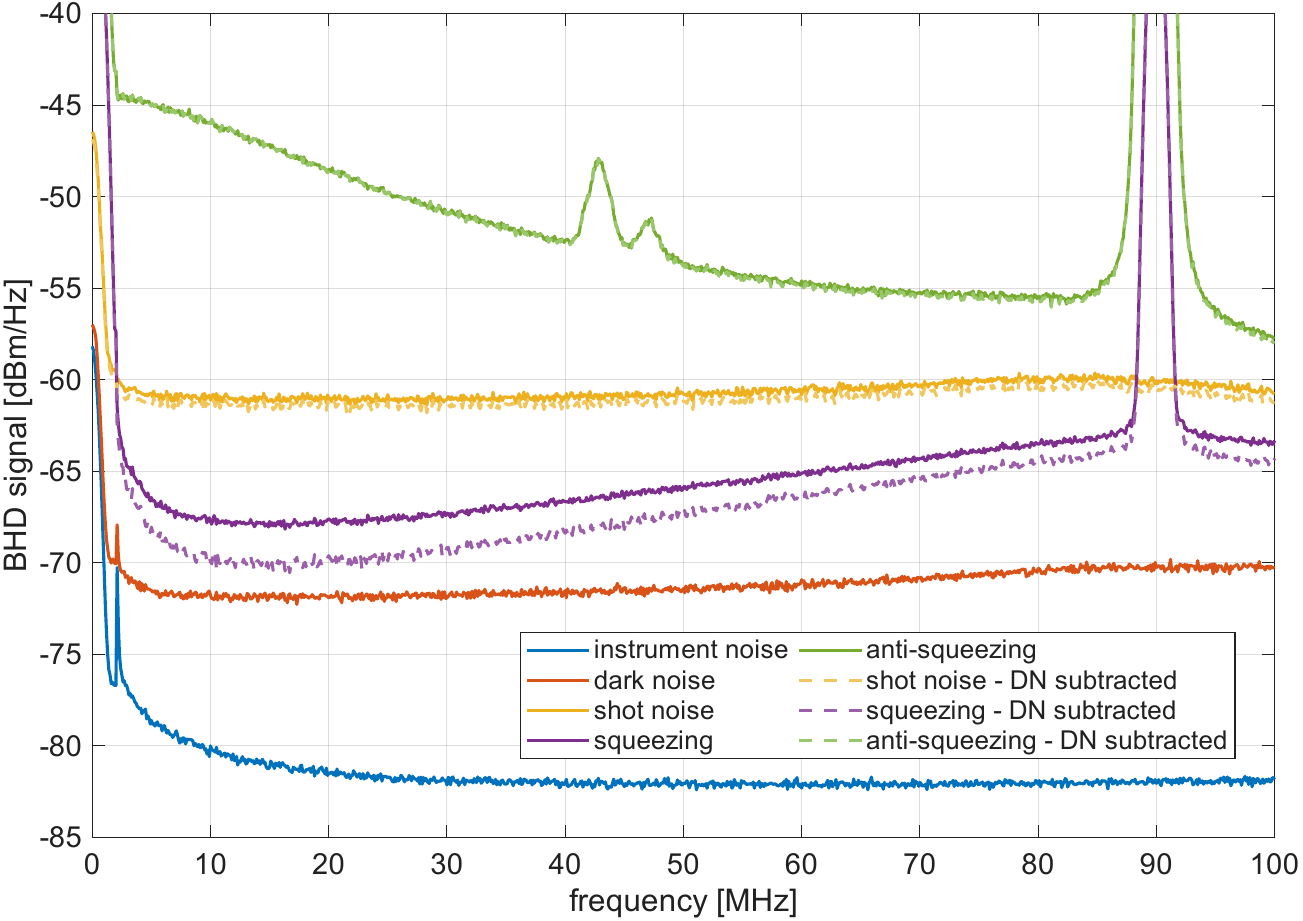}
\caption{\label{fig:spectra} BHD signals measured with the spectrum analyser using an RBW of 1 MHz and averaging over 1000 traces. The dashed traces were obtained by subtracting the dark noise from the shot noise traces. The peak at 89.9 MHz is the phase modulation imprinted by the EOM inside the SLS as part of SHG and OPA length stabilisation control loops. The peaks at 42.3 MHz and 47.6 MHz are aliasing artifacts of the third and fourth harmonics of the 89.9 MHz peak, respectively, given the Nyquist frequency of 156 MHz.
These features were still visible due to the finite attenuation of the anti-aliasing filters (ZX75LP-137-S+ from Mini-circuits), which have a 137 MHz cutoff frequency.}}
\end{figure}
\indent The BHD signals were recorded with a spectrum analyser using a span of 100\,MHz and resolution bandwidth (RBW) of 1\,MHz, as shown in Fig.~\ref{fig:spectra}. The instrument noise was obtained by unplugging the spectrum analyser from the digital subtractor, the dark noise was obtained by blocking both the squeezed light and the LO path, the coherent shot noise was obtained by letting only the LO reach the PDs.\\
\indent Two strategies were employed for readout phase control in the perpendicular quadratures: in the squeezing configuration, the error signal was obtained by demodulating at 89.9 MHz (RF control); in the anti-squeezing configuration, the error signal was derived from the DC signal of the PDs (DC control). To improve the SNR and enhance the robustness against power fluctuations, the error signal for both control loops was obtained from the subtraction of the two PDs, rather than from a single PD.
The actuator for the readout phase control was a piezo-mounted mirror in the LO path.\\
\indent The effectiveness of this control scheme is demonstrated in Fig.~\ref{fig:spectrogram}, which shows the long-term operation of the BHD. For the first 30 seconds, a beam dump blocked the squeezed light field. After its removal, the BHD phase was locked to the squeezed quadrature. Over the subsequent one-hour observation, the system remained stably locked, with the exception of a few short glitches of excess noise that did not disrupt the lock. 
The beam dump was then reinserted to intentionally terminate the lock—although the system could have remained locked for a longer duration. These results confirm the suitability of the setup for stable long-term operation, not only in terms of readout angle control, but also with respect to the overall stability of all sub-systems (SHG, OPA, MZ).

\section{Influence of Imperfections}\label{sec:imperfections}
Squeezed states of light are highly susceptible to degradation due to optical loss and phase noise. To assess the influence of imperfections, we compared our results with a theoretical model.
For an OPA cavity operating below threshold, the variances of the squeezed ($V_-$) and anti-squeezed ($V_+$) quadratures of the state reaching the readout scheme (in this case, the BHD) in the absence of phase noise are given by \cite{polzik1992atomic}:
\begin{equation}\label{eq:loss}
    V^\eta_\pm=1\pm\eta\frac{4\sqrt{P/P_\mathrm{thres}}}{(1\mp\sqrt{P/P_\mathrm{thres}})^2+4\left(\frac{2\pi f}{\gamma}\right)^2}\;,
\end{equation}
where $\eta$ is the total detection efficiency, $P$ is the pump power, $P_\mathrm{thres}$ is the OPA threshold power, $f$ is the sideband frequency of the measurement, and $\gamma=c(T+L)l_\mathrm{opt}^{-1}$ is the cavity decay rate. In the last expression, $T$ is the power transmissivity of the output coupler, $L$ is the round-trip loss inside the OPA, $c$ is the speed of light, and $l_\mathrm{opt}$ is the optical round-trip length of the OPA. The cavity linewidth, expressed in terms of full width at half maximum, is given by $\gamma/(2\pi)$ \cite{mckenzie2008squeezing}. The total detection efficiency can be factorised into two contributions: $\eta=\eta_\mathrm{esc}\eta_\mathrm{det}$, where $\eta_\mathrm{esc}=T/(T+L)$ is the OPA escape efficiency, which depends only on the OPA parameters. The detection efficiency $\eta_\mathrm{det}$ reflects how effectively the SLS is integrated into the readout scheme; it accounts for factors such as propagation loss, mode overlap between the squeezed field and the LO, and the quantum efficiency of the PDs.\\
\indent The effect of phase fluctuations between LO and squeezed field is to mix squeezed and anti-squeezed quadratures according to \cite{aoki2006squeezing}
\begin{equation}\label{eq:phase_noise}
    V^{\eta,\theta}_\pm=V^\eta_\pm\cos^2\theta+V^\eta_{\mp}\sin^2\theta\;,
\end{equation}
where $\theta$ is the RMS value of the phase noise, assumed to be Gaussian.\\
\indent The parameters in Eqs.~\ref{eq:loss} and \ref{eq:phase_noise}—$\eta$, $\theta$, $P_{\mathrm{thres}}$, and $\gamma$—were estimated as follows.\\
\indent First, the RMS phase noise was estimated from a calibrated spectrum of the in-loop phase control error signal.
Two different error signals were used for phase control of the two quadratures: the DC control error signal had a peak-to-peak amplitude approximately 16 times smaller than that of the RF control error signal. The loop gain was adjusted accordingly to maintain the same unity gain frequency (UGF) for both control loops. The UGF was 75\,Hz, limited by a mechanical resonance at 270\,Hz, likely originating from the mount of the piezoelectric actuator. The resulting RMS phase noise was $\theta=16\,\mathrm{mrad}$ for both quadratures.\\
\indent Second, $P_{\mathrm{thres}}$ and $\eta$ were obtained by fitting the measured squeezing and anti-squeezing levels as a function of pump power at several sideband frequencies (from 16 MHz to 40 MHz at steps of 1 MHz) and averaging the resulting parameters, yielding $P_{\mathrm{thres}} = 163(3)\,\mathrm{mW}$ and $\eta=\,90.6(4)\,\%$. This result was obtained after removing the dark noise from the shot noise traces. A fit of the traces obtained without removing the dark noise yielded $\eta=82.8(5)\,\%$.\\
\indent Finally, the remaining parameter $\gamma$ was obtained by fitting
the measured squeezing and anti-squeezing levels as a function of frequency at several pump powers (from 40 mW to 90 mW at steps of 10 mW), with $P_{\mathrm{thres}}$, $\eta$, and $\theta$ fixed to their previously estimated values. The final value of $\gamma$ was obtained by averaging the results of these fits. The fit yielded $\gamma/(2\pi) = 138(2)\,\mathrm{MHz}$.\\
\indent Note that the first fit (squeezing versus pump power) also depends weakly on the cavity linewidth $\gamma/(2\pi)$, which was initially fixed to a reasonable guess value. Conversely, the second fit (squeezing versus frequency) depends on $P_{\mathrm{thres}}$ and $\eta$, previously obtained from the first fit. To account for this mutual dependence, the two fitting procedures were iterated: the output of one fit was used as input to the other until the parameter values converged. The final values reported above were obtained after this self-consistent iteration.\\
\begin{figure}[h!]
\center{
\includegraphics[width=.7\textwidth]{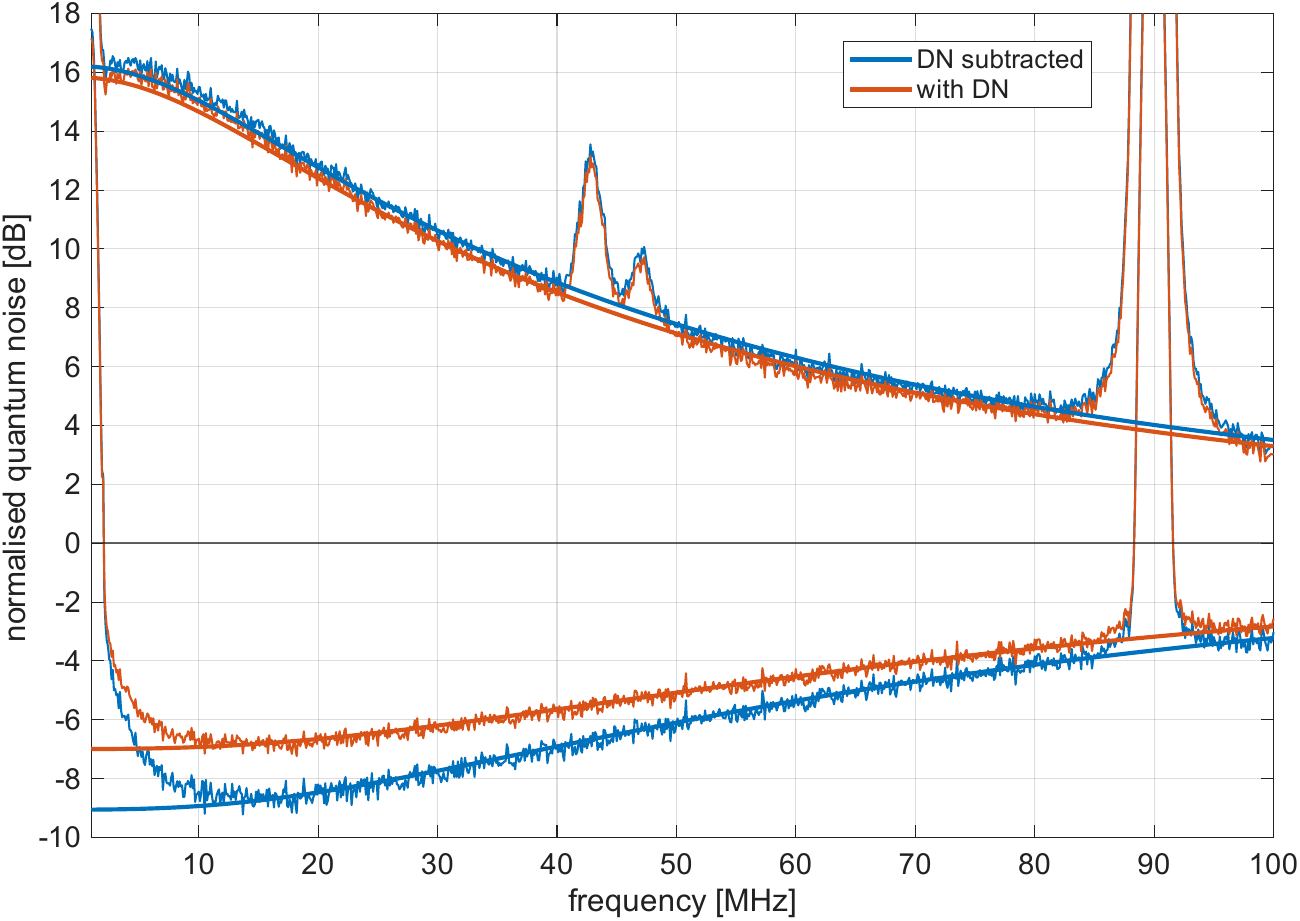}
\caption{\label{fig:sqz_asqz_spectra} Normalised quantum noise spectra. Both cases—with and without dark noise subtraction—are shown, along with the fitted model. The parameters used for the model were $\eta_{\mathrm{DN\,subtracted}}=90.6\,\%$, $\eta_{\mathrm{with\,DN}}=82.8\,\%$, $\theta=16\,\mathrm{mrad}$, $P=90\,\mathrm{mW}$, $P_\mathrm{thres}=163\,\mathrm{mW}$, $\gamma/(2\pi)=138\,\mathrm{MHz}$.}}
\end{figure}
\indent Fig.~\ref{fig:sqz_asqz_spectra} shows the best result obtained with the BHD measurement, along with the model prediction based on the parameters obtained above. The traces were produced by dividing the `squeezing' and `anti-squeezing' spectra from Fig.~\ref{fig:spectra} by the `shot noise' trace. The highest directly observed squeezing level was 6.8 dB at a sideband frequency of 17 MHz. After subtracting the dark noise contribution, the inferred squeezing level increased to 8.6 dB at the same frequency.\\
\indent In the following, we consider the possible contributions to the total optical loss. A summary of the estimated loss budget is provided in Table~\ref{tab:loss_budget}.\\
\begin{table}[h!]
    \centering
    \begin{tabular}{l l}
        Source &  Loss (\%)\\
        \hline\hline
        Dark-noise equivalent loss & 8.6\\
        Propagation loss& 2.70(3)\\
        Homodyne efficiency & 2.6(9)\\
        PD quantum efficiency & 1\\
        OPA escape efficiency& 1\\
        99:1 pick-off after OPA & 1\\
        PD Reflection & 0.25(2)\\
        BS splitting ratio & 0.004\\
        Unknown source& 1.2\\
        \hline
        Total &17.2\\
    \end{tabular}
    \caption{Optical loss contributions. The sum of the known losses does not fully account for the estimated total optical loss. The remaining discrepancy is included as `unknown source' to reflect uncharacterised losses in the system.}
    \label{tab:loss_budget}
\end{table}
\indent The dark-noise equivalent loss \cite{appel2007electronic} was $8.6\,\%$ for a dark-noise clearance of $10.6\,\mathrm{dB}$ with a LO power of 17.2 mW. The available LO power was limited because the initial laser output (430 mW) had to be split between the BHD LO (20 mW) and the SLS input (410 mW). Consequently, only a small fraction of power could be allocated to the LO. However, in the final implementation of the SLS in the QUEST experiment, the power available at the readout PD will be 40 mW, enabling a dark-noise clearance of 14 dB and a corresponding equivalent loss of 4\,\%.
The relatively high dark noise is not easily reduced, except by using a self-subtraction PD design, as typically employed in BHD \cite{vahlbruch2008squeezed}. We chose not to adopt this approach, since in the main application of the SLS, BHD is not foreseen in the near future. 
Instead, we used the same PDs that will be employed in the QUEST experiment, each equipped with its own dedicated electronics.\\
\indent The propagation loss refers to losses from optics—due to absorption or reflection—along the path of the squeezed field. In our setup, the propagation loss is dominated by the lenses: two in the mode-matching telescope and one in front of each PD. We estimated an average of 0.4\,\% of residual reflection per lens surface.\\
\indent The homodyne efficiency quantifies the mode overlap between the two interfering beams (LO and squeezed light), and is given by $\eta_\mathrm{HD} = \mathrm{VIS}^2$~\cite{wilken2024high}, where the visibility ($\mathrm{VIS}$) is measured using two beams of equal power.
Since the two interfering beams had different powers, the homodyne efficiency was estimated using
    \begin{equation}
        \eta_\mathrm{HD}=\left(\frac{P_\mathrm{min}-P_\mathrm{max}}{P_\mathrm{min}+P_\mathrm{max}}/\frac{2\sqrt{P_\mathrm{LO}P_\mathrm{BAB}}}{P_\mathrm{LO}+P_\mathrm{BAB}}\right)^2\;,
    \end{equation}
where $P_\mathrm{min}$ and $P_\mathrm{max}$ are the power minima and maxima of the interfered beams, respectively, $P_\mathrm{LO}=17.2\,\mathrm{mW}$ is the LO power, and $P_\mathrm{BAB}=\SI{31}{\micro\watt}$ is the BAB power. Despite the fact that the mode-matching to the reference cavity was 99.8\% for the BAB and 99.7\% for the LO, we obtained a homodyne efficiency of $\eta_\mathrm{HD}=97.4(9)\,\%$, probably due to residual misalignment and polarisation mismatch.\\
\indent The non-perfect splitting ratio of the BS is equivalent to a loss of $1-\eta_\mathrm{BS}=1-4R(1-R)$ \cite{wilken2024high}, where $R$ is the BS reflectivity. In our setup, we had $R=50.3\,\%$, meaning an equivalent loss of 0.004\,\%.\\
\indent The common-mode rejection ratio (CMRR) \cite{wilken2024high} represents the capability of a BHD to eliminate noise common to both inputs, particularly important at the technical noise limited low frequency. We did not have a setup to measure the CMRR in a frequency-dependent way, but we could get an estimation of the CMRR at the relaxation oscillation frequency of the laser (450 kHz) by comparing the peak height at BHD normal operation and by blocking one PD: the CMRR at $450\,\mathrm{kHz}$ was 45\,dB. 
The relatively low CMRR at low frequencies—and the resulting high technical noise below approximately $10\,\mathrm{MHz}$—was due to the fact that all measurements were performed with the laser noise eater not engaged, as it was unavailable due to a technical issue.

\section{Conclusions}
In this work, we directly observed 6.8 dB of squeezing at a sideband frequency of 17 MHz using a BHD, and maintained at least 3 dB of squeezing across a 100 MHz bandwidth. After subtracting the dark noise contribution, the inferred squeezing level increased to 8.6 dB.
The observed limitation in squeezing level is primarily attributed to detection loss, which we addressed by developing a comprehensive loss budget.  Several limitations encountered in the current setup are expected to be mitigated in the forthcoming QUEST setup. In particular, increasing the LO power from 17.2\,mW to 40\,mW will improve the dark-noise clearance, while the faster acquisition system will extend the detection bandwidth.\\
\indent A comparison of our results with existing literature shows that the SLS for the QUEST experiment has the broadest linewidth, among those suitable for long-term operation. The stability of the system over extended timescales was demonstrated in this work.\\
\indent Having demonstrated a parametric gain of at least 16 dB, and knowing the intrinsic loss of our squeezed light source to be 2\,\%—set by the OPA escape efficiency and the 1\,\% pick-off used for the OPA longitudinal control—we can use these results to inform the design of the squeezed light injection path for the QUEST experiment, particularly regarding tolerable optical loss. To achieve 6\,dB of detected squeezing starting from a pure 16 dB squeezed state, and assuming an RMS phase noise of 10 mrad, we estimate that up to 21\,\% additional optical loss (on top of the intrinsic 2\,\%) can be tolerated.

\section {Funding}
Funded by the Deutsche Forschungsgemeinschaft (DFG, German Research Foundation) under Germany’sExcellenceStrategy – EXC-2123 QuantumFrontiers – 390837967.

\printbibliography

@article{Ast:12,
author = {Stefan Ast and Aiko Samblowski and Moritz Mehmet and Sebastian Steinlechner and Tobias Eberle and Roman Schnabel},
journal = {Optics Letters},
number = {12},
pages = {2367--2369},
publisher = {Optica Publishing Group},
title = {Continuous-wave nonclassical light with gigahertz squeezing bandwidth},
volume = {37},
year = {2012}
}

@article{wilken2024broadband,
  title={Broadband detection of 18 teeth in an 11-dB squeezing comb},
  author={Wilken, Dennis and Junker, Jonas and Heurs, Mich{\`e}le},
  journal={Physical Review Applied},
  volume={21},
  number={3},
  pages={031002},
  year={2024},
  publisher={APS}
}

@article{drever1983laser,
  title={Laser phase and frequency stabilization using an optical resonator},
  author={Drever, Ronald WP and Hall, John L and Kowalski, Frank V and Hough, James and Ford, GM and Munley, AJ and Ward, Hywel},
  journal={Applied Physics B},
  volume={31},
  pages={97--105},
  year={1983},
  publisher={Springer}
}

@article{ast2013high,
  title={High-bandwidth squeezed light at 1550 nm from a compact monolithic PPKTP cavity},
  author={Ast, Stefan and Mehmet, Moritz and Schnabel, Roman},
  journal={Optics Express},
  volume={21},
  number={11},
  pages={13572--13579},
  year={2013},
  publisher={Optical Society of America}
}

@article{serikawa2016creation,
  title={Creation and measurement of broadband squeezed vacuum from a ring optical parametric oscillator},
  author={Serikawa, Takahiro and Yoshikawa, Jun-ichi and Makino, Kenzo and Frusawa, Akira},
  journal={Optics Express},
  volume={24},
  number={25},
  pages={28383--28391},
  year={2016},
  publisher={Optical Society of America}
}

@article{mehmet2010observation,
  title={Observation of squeezed states with strong photon-number oscillations},
  author={Mehmet, Moritz and Vahlbruch, Henning and Lastzka, Nico and Danzmann, Karsten and Schnabel, Roman},
  journal={Physical Review A—Atomic, Molecular, and Optical Physics},
  volume={81},
  number={1},
  pages={013814},
  year={2010},
  publisher={APS}
}

@article{vahlbruch2016detection,
  title={Detection of 15 dB squeezed states of light and their application for the absolute calibration of photoelectric quantum efficiency},
  author={Vahlbruch, Henning and Mehmet, Moritz and Danzmann, Karsten and Schnabel, Roman},
  journal={Physical Review Letters},
  volume={117},
  number={11},
  pages={110801},
  year={2016},
  publisher={APS}
}

@article{meylahn2022squeezed,
  title={Squeezed states of light for future gravitational wave detectors at a wavelength of 1550 nm},
  author={Meylahn, Fabian and Willke, Benno and Vahlbruch, Henning},
  journal={Physical Review Letters},
  volume={129},
  number={12},
  pages={121103},
  year={2022},
  publisher={APS}
}

@article{lough2021first,
  title={First demonstration of 6 dB quantum noise reduction in a kilometer scale gravitational wave observatory},
  author={Lough, James and Schreiber, Emil and Bergamin, Fabio and Grote, Hartmut and Mehmet, Moritz and Vahlbruch, Henning and Affeldt, Christoph and Brinkmann, Marc and Bisht, Aparna and Kringel, Volker and others},
  journal={Physical Review Letters},
  volume={126},
  number={4},
  pages={041102},
  year={2021},
  publisher={APS}
}

@article{heinze2022observation,
  title={Observation of squeezed states of light in higher-order Hermite-Gaussian modes with a quantum noise reduction of up to 10 dB},
  author={Heinze, Joscha and Willke, Benno and Vahlbruch, Henning},
  journal={Physical Review Letters},
  volume={128},
  number={8},
  pages={083606},
  year={2022},
  publisher={APS}
}

@article{fejer2002quasi,
  title={Quasi-phase-matched second harmonic generation: tuning and tolerances},
  author={Fejer, Martin M and Magel, GA and Jundt, Dieter H and Byer, Robert L},
  journal={IEEE Journal of Quantum Electronics},
  volume={28},
  number={11},
  pages={2631--2654},
  year={2002},
  publisher={IEEE}
}

@article{mehmet2011squeezed,
  title={Squeezed light at 1550 nm with a quantum noise reduction of 12.3 dB},
  author={Mehmet, Moritz and Ast, Stefan and Eberle, Tobias and Steinlechner, Sebastian and Vahlbruch, Henning and Schnabel, Roman},
  journal={Optics Express},
  volume={19},
  number={25},
  pages={25763--25772},
  year={2011},
  publisher={Optical Society of America}
}

@article{patra2024direct,
  title={Broadband limits on stochastic length fluctuations from a pair of table-top interferometers},
  author={Patra, Abhinav and Aiello, Lorenzo and Ejlli, Aldo and Griffiths, William L and James, Alasdair L and Kuntimaddi, Nikitha and Kwon, Ohkyung and Schwartz, Eyal and Vahlbruch, Henning and Vermeulen, Sander M and others},
  journal={arXiv preprint arXiv:2410.09175},
  year={2024}
}

@article{vermeulen2021experiment,
  title={An experiment for observing quantum gravity phenomena using twin table-top 3D interferometers},
  author={Vermeulen, Sander M and Aiello, Lorenzo and Ejlli, Aldo and Griffiths, William L and James, Alasdair L and Dooley, Katherine L and Grote, Hartmut},
  journal={Classical and Quantum Gravity},
  volume={38},
  number={8},
  pages={085008},
  year={2021},
  publisher={IOP Publishing}
}

@phdthesis{griffiths2023enhancing,
  title={Enhancing the QUEST experiment with a coin sized Output Mode Cleaner for improved sensitivity to Quantum Gravity signatures},
  author={Griffiths, William},
  year={2023},
  school={Cardiff University}
}

@article{tohermes2024directly,
  title={Directly measured squeeze factors over GHz bandwidth from monolithic ppKTP resonators},
  author={Tohermes, Benedict and Verclas, Sophie and Schnabel, Roman},
  journal={arXiv preprint arXiv:2412.03221},
  year={2024}
}

@article{kashiwazaki2023over,
  title={Over-8-dB squeezed light generation by a broadband waveguide optical parametric amplifier toward fault-tolerant ultra-fast quantum computers},
  author={Kashiwazaki, Takahiro and Yamashima, Taichi and Enbutsu, Koji and Kazama, Takushi and Inoue, Asuka and Fukui, Kosuke and Endo, Mamoru and Umeki, Takeshi and Furusawa, Akira},
  journal={Applied Physics Letters},
  volume={122},
  number={23},
  year={2023},
  publisher={AIP Publishing}
}

@article{kashiwazaki2021fabrication,
  title={Fabrication of low-loss quasi-single-mode PPLN waveguide and its application to a modularized broadband high-level squeezer},
  author={Kashiwazaki, Takahiro and Yamashima, Taichi and Takanashi, Naoto and Inoue, Asuka and Umeki, Takeshi and Furusawa, Akira},
  journal={Applied Physics Letters},
  volume={119},
  number={25},
  year={2021},
  publisher={AIP Publishing}
}

@article{wu1987squeezed,
  title={Squeezed states of light from an optical parametric oscillator},
  author={Wu, Ling-An and Xiao, Min and Kimble, HJ},
  journal={Journal of the Optical Society of America B},
  volume={4},
  number={10},
  pages={1465--1475},
  year={1987},
  publisher={Optical Society of America}
}

@article{pysher2009broadband,
  title={Broadband amplitude squeezing in a periodically poled $\mathrm{KTiOPO_4}$ waveguide},
  author={Pysher, Matthew and Bloomer, Russell and Kaleva, Christopher M and Roberts, Tony D and Battle, Philip and Pfister, Olivier},
  journal={Optics Letters},
  volume={34},
  number={3},
  pages={256--258},
  year={2009},
  publisher={Optical Society of America}
}

@article{caves1981quantum,
  title={Quantum-mechanical noise in an interferometer},
  author={Caves, Carlton M},
  journal={Physical Review D},
  volume={23},
  number={8},
  pages={1693},
  year={1981},
  publisher={APS}
}

@article{ralph1999continuous,
  title={Continuous variable quantum cryptography},
  author={Ralph, Timothy C},
  journal={Physical Review A},
  volume={61},
  number={1},
  pages={010303},
  year={1999},
  publisher={APS}
}

@article{hillery2000quantum,
  title={Quantum cryptography with squeezed states},
  author={Hillery, Mark},
  journal={Physical Review A},
  volume={61},
  number={2},
  pages={022309},
  year={2000},
  publisher={APS}
}

@article{purdy2013strong,
  title={Strong optomechanical squeezing of light},
  author={Purdy, Thomas P and Yu, P-L and Peterson, Robert W and Kampel, Nir S and Regal, Cindy A},
  journal={Physical Review X},
  volume={3},
  number={3},
  pages={031012},
  year={2013},
  publisher={APS}
}

@article{shelby1986broad,
  title={Broad-band parametric deamplification of quantum noise in an optical fiber},
  author={Shelby, Robert M and Levenson, Marc D and Perlmutter, Stephen H and DeVoe, Ralph G and Walls, Daniel F},
  journal={Physical Review Letters},
  volume={57},
  number={6},
  pages={691},
  year={1986},
  publisher={APS}
}

@article{slusher1985observation,
  title={Observation of squeezed states generated by four-wave mixing in an optical cavity},
  author={Slusher, R E and Hollberg, LW and Yurke, Bernard and Mertz, JC and Valley, JF},
  journal={Physical Review Letters},
  volume={55},
  number={22},
  pages={2409},
  year={1985},
  publisher={APS}
}

@phdthesis{hage2010purification,
  title={Purification and Distillation of Continuous Variable Entanglement},
  author={Hage, Boris},
  year={2010},
  school={Leibniz Universit{\"a}t Hannover}
}

@article{wu1986generation,
  title={Generation of squeezed states by parametric down conversion},
  author={Wu, Ling-An and Kimble, HJ and Hall, JL and Wu, Huifa},
  journal={Physical Review Letters},
  volume={57},
  number={20},
  pages={2520},
  year={1986},
  publisher={APS}
}

@phdthesis{wilken2024high,
  title={A high-frequency squeezing comb-generation, detection \& characterisation},
  author={Wilken, Dennis Max},
  year={2024},
  school={Leibniz Universit{\"a}t Hannover}
}

@article{appel2007electronic,
  title={Electronic noise in optical homodyne tomography},
  author={Appel, J{\"u}rgen and Hoffman, Dallas and Figueroa, Eden and Lvovsky, AI},
  journal={Physical Review A—Atomic, Molecular, and Optical Physics},
  volume={75},
  number={3},
  pages={035802},
  year={2007},
  publisher={APS}
}

@article{vahlbruch2010geo,
  title={The GEO 600 squeezed light source},
  author={Vahlbruch, Henning and Khalaidovski, Alexander and Lastzka, Nico and Gr{\"a}f, Christian and Danzmann, Karsten and Schnabel, Roman},
  journal={Classical and Quantum Gravity},
  volume={27},
  number={8},
  pages={084027},
  year={2010},
  publisher={IOP Publishing}
}

@article{mehmet2018high,
  title={High-efficiency squeezed light generation for gravitational wave detectors},
  author={Mehmet, Moritz and Vahlbruch, Henning},
  journal={Classical and Quantum Gravity},
  volume={36},
  number={1},
  pages={015014},
  year={2018},
  publisher={IOP Publishing}
}

@article{schonbeck201713,
  title={13 dB squeezed vacuum states at 1550 nm from 12 mW external pump power at 775 nm},
  author={Sch{\"o}nbeck, Axel and Thies, Fabian and Schnabel, Roman},
  journal={Optics Letters},
  volume={43},
  number={1},
  pages={110--113},
  year={2017},
  publisher={Optical Society of America}
}

@article{darsow2021squeezed,
  title={Squeezed light at 2128 nm for future gravitational-wave observatories},
  author={Darsow-Fromm, Christian and Gurs, Julian and Schnabel, Roman and Steinlechner, Sebastian},
  journal={Optics Letters},
  volume={46},
  number={23},
  pages={5850--5853},
  year={2021},
  publisher={Optical Society of America}
}

@article{tse2019quantum,
  title={Quantum-enhanced advanced LIGO detectors in the era of gravitational-wave astronomy},
  author={Tse, Maggie and Yu, Haocun and Kijbunchoo, Nutsinee and Fernandez-Galiana, A and Dupej, P and Barsotti, L and Blair, CD and Brown, DD and Dwyer, SE ea and Effler, A and others},
  journal={Physical Review Letters},
  volume={123},
  number={23},
  pages={231107},
  year={2019},
  publisher={APS}
}

@article{acernese2019increasing,
  title={Increasing the astrophysical reach of the Advanced Virgo detector via the application of squeezed vacuum states of light},
  author={Acernese, Fausto and Agathos, M and Aiello, L and Allocca, A and Amato, A and Ansoldi, S and Antier, S and Ar{\`e}ne, M and Arnaud, N and Ascenzi, S and others},
  journal={Physical Review Letters},
  volume={123},
  number={23},
  pages={231108},
  year={2019},
  publisher={APS}
}

@phdthesis{vahlbruch2008squeezed,
  title={Squeezed light for gravitational wave astronomy},
  author={Vahlbruch, Henning},
  year={2008},
  school={Leibniz Universit{\"a}t Hannover}
}

@article{mehmet2020squeezed,
  title={The squeezed light source for the Advanced Virgo detector in the observation run O3},
  author={Mehmet, Moritz and Vahlbruch, Henning},
  journal={Galaxies},
  volume={8},
  number={4},
  pages={79},
  year={2020},
  publisher={MDPI}
}

@article{vahlbruch2006coherent,
  title={Coherent control of vacuum squeezing in the gravitational-wave detection band},
  author={Vahlbruch, Henning and Chelkowski, Simon and Hage, Boris and Franzen, Alexander and Danzmann, Karsten and Schnabel, Roman},
  journal={Physical Review Letters},
  volume={97},
  number={1},
  pages={011101},
  year={2006},
  publisher={APS}
}

@article{polzik1992atomic,
  title={Atomic spectroscopy with squeezed light for sensitivity beyond the vacuum-state limit},
  author={Polzik, ES and Carri, J and Kimble, HJ},
  journal={Applied Physics B},
  volume={55},
  number={3},
  pages={279--290},
  year={1992},
  publisher={Springer}
}

@article{aoki2006squeezing,
  title={Squeezing at 946nm with periodically poled KTiOPO4},
  author={Aoki, Takao and Takahashi, Go and Furusawa, Akira},
  journal={Optics Express},
  volume={14},
  number={15},
  pages={6930--6935},
  year={2006},
  publisher={Optical Society of America}
}

@article{schnabel2017squeezed,
  title={Squeezed states of light and their applications in laser interferometers},
  author={Schnabel, Roman},
  journal={Physics Reports},
  volume={684},
  pages={1--51},
  year={2017},
  publisher={Elsevier}
}

@article{dwyer2022squeezing,
  title={Squeezing in gravitational wave detectors},
  author={Dwyer, Sheila E and Mansell, Georgia L and McCuller, Lee},
  journal={Galaxies},
  volume={10},
  number={2},
  pages={46},
  year={2022},
  publisher={MDPI}
}

@phdthesis{mckenzie2008squeezing,
  title={Squeezing in the audio gravitational wave detection band},
  author={McKenzie, Kirk},
  year={2008},
  school={Australian National University}
}

@phdthesis{buchler2001electro,
  title={Electro-optic control of quantum measurements},
  author={Buchler, Benjamin},
  year={2001},
  school={Australian National University}
}

@phdthesis{eberle2013realization,
  title={Realization of finite-size quantum key distribution based on Einstein-Podolsky-Rosen entangled light},
  author={Eberle, Tobias},
  year={2013},
  school={Leibniz Universit{\"a}t Hannover}
}

\end{document}